%% file: template.tex
\documentclass[a4paper]{jpconf}
\usepackage{graphicx}
\usepackage{upgreek}
\usepackage{xcolor}
\usepackage{amsmath, amsfonts, amssymb}
\usepackage{ptdr-definitions}
\usepackage[caption=false]{subfig}
\usepackage{xspace}

\makeatletter
\DeclareRobustCommand*{\bfseries}{\not@math@alphabet\bfseries\mathbf\fontseries\bfdefault\selectfont\boldmath}
\makeatother

\begin{document}
\title{Experimental searches for tHq}

\author{Christian B\"oser on behalf of the ATLAS and CMS collaborations}
\address{Institut f\"ur Experimentelle Kernphysik, Karlsruhe, Germany}

\ead{christian.boeser@kit.edu}

\begin{abstract}
The Yukawa coupling of the Higgs boson to top quarks $\Yt$ is of particular interest as the top quark plays a special role in the electroweak symmetry breaking mechanism. Any deviation from its predicted value could hint at new physics. For this investigation the associated production of a Higgs boson and a single top quark ($\tHq$) is a promising channel. Due to interference terms in the production amplitude the cross section $\sigma_\tHq$ is not only sensitive to the magnitude of $\Yt$ but also to its sign making this channel special. In addition the branching ratio BR($\Hgg$) is sensitive to sign and magnitude of $\Yt$. The ATLAS and CMS collaborations exploit both effects to set upper limits on $\sigma_\tHq$ and the deviation of $\Yt$ by applying inherently different approaches. The most recent results from both collaborations using the full data provided by the LHC are presented.
%
%
\end{abstract}

\section{Introduction}
The discovery of a Higgs boson by the ATLAS~\cite{ATLASdetector} and CMS collaborations~\cite{CMSdetector} in 2012~\cite{ATLASHiggsObs,CMS-PAPERS-HIG-12-036} was a major success. Since then many analyses tried to reveal its properties and to answer the question, whether it is in fact the Higgs boson predicted by the standard model (SM). However, more precise measurements are needed to rule out theories that forecast a differing appearance of the Higgs boson. 

\begin{figure}[hbtp]
\centering
\subfloat{\includegraphics[width=0.23\textwidth]{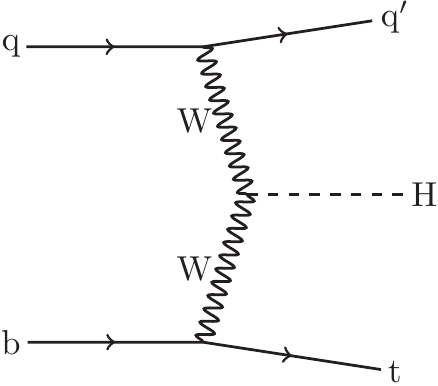}}
\hspace{3cm}
 \subfloat{\includegraphics[width=0.23\textwidth]{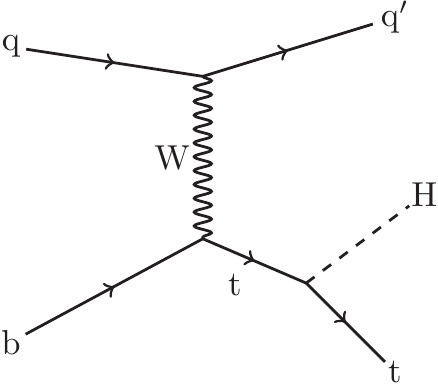}} \\
\caption{Representative Feynman diagrams for the $t$-channel single top quark production in association with a Higgs boson \cPqb\cPq $\rightarrow \tHq^\prime$. The destructive interference between these two processes leads to a small cross section for $\tHq$ production in the SM.}
\label{fig:feynman}
\end{figure}

Of particular interest is the coupling $\Yt^{\text{SM}}$ of the Higgs boson to the heaviest fermion, the top quark, since it gives insights into understanding the electroweak symmetry breaking. Deviations to this coupling $\Yt = \kt\Yt^{\text{SM}}$ are predicted by theoretical modifications beyond the SM, so it is important to determine $\kt$ as accurate as possible. Earlier this year, the ATLAS and CMS collaborations updated the constraints on $\kt$ using all available Higgs boson analyses~\cite{ATLAS-CONF-2014-009, CMS-PAS-HIG-14-009}. While both experiments favor the SM scenario with $\kt = 1$, the anomalous case of a flipped sign $\kt = -1$ is still tolerated when contributions beyond the SM are allowed~\cite{Ellis}.
The reason for that is the sign-insensitive dependence of the dominant Higgs boson production modes. Also the production of a Higgs boson in association with a top quark pair is only sensitive to the square of $\kt$.  The only sign-sensitive constraints come from the search for $\Hgg$ decays. The branching ratio BR($\Hgg$) is affected by the interference of contributions from top quark and W bosons in the decay loop, so $\kt$ can be directly probed. For example a 2.4 times enhanced branching ratio compared to the SM is expected for the anomalous scenario $\kt = -1$.
An even more sensitive channel is constituted by the $t$-channel single top quark production in association with a Higgs boson (hereafter $\tHq$). Figure~\ref{fig:feynman} shows the two dominant Feynman diagrams which interfere with each other destructively in the SM. This leads to a small cross section of 18.3\,fb at a center-of-mass energy of 8\,TeV~\cite{Maltoni,Farina}. Yet an enhanced cross section for $\tHq$ production is predicted by any variation of $\kt$, for example expected from CP violation~\cite{Chang:2014rfa}. The direct search for $\tHq$ production assuming an anomalous of coupling $\kt = -1$, that leads to a 14 times enhanced cross section, represents a nice backdoor to make the most of the data recorded so far at the LHC.

In the following sections first the two available direct searches for $\tHq$ production with $\Hgg$ and $\Hbb$ decays assuming an anomalous coupling from the CMS collaboration are presented. Both analyses use the full 8 TeV dataset recorded by the CMS detector corresponding to an integrated luminosity of $20\fbinv$. Additionally shown is the analysis from the ATLAS collaboration setting constraints on $\kt$ via the search for $\ttH$ production with $\Hgg$. The whole dataset recorded with the ATLAS experiment corresponding to an integrated luminosity of $4.5\fbinv$ (7 TeV) and $20.3\fbinv$ (8 TeV) is used. 

\section{Direct search for $\tHq$ and $\Hgg$ with the CMS experiment}
\begin{figure}[hbtp]
\begin{center}
\begin{minipage}[t]{0.475\textwidth}
\centering
\includegraphics[width=0.83\textwidth]{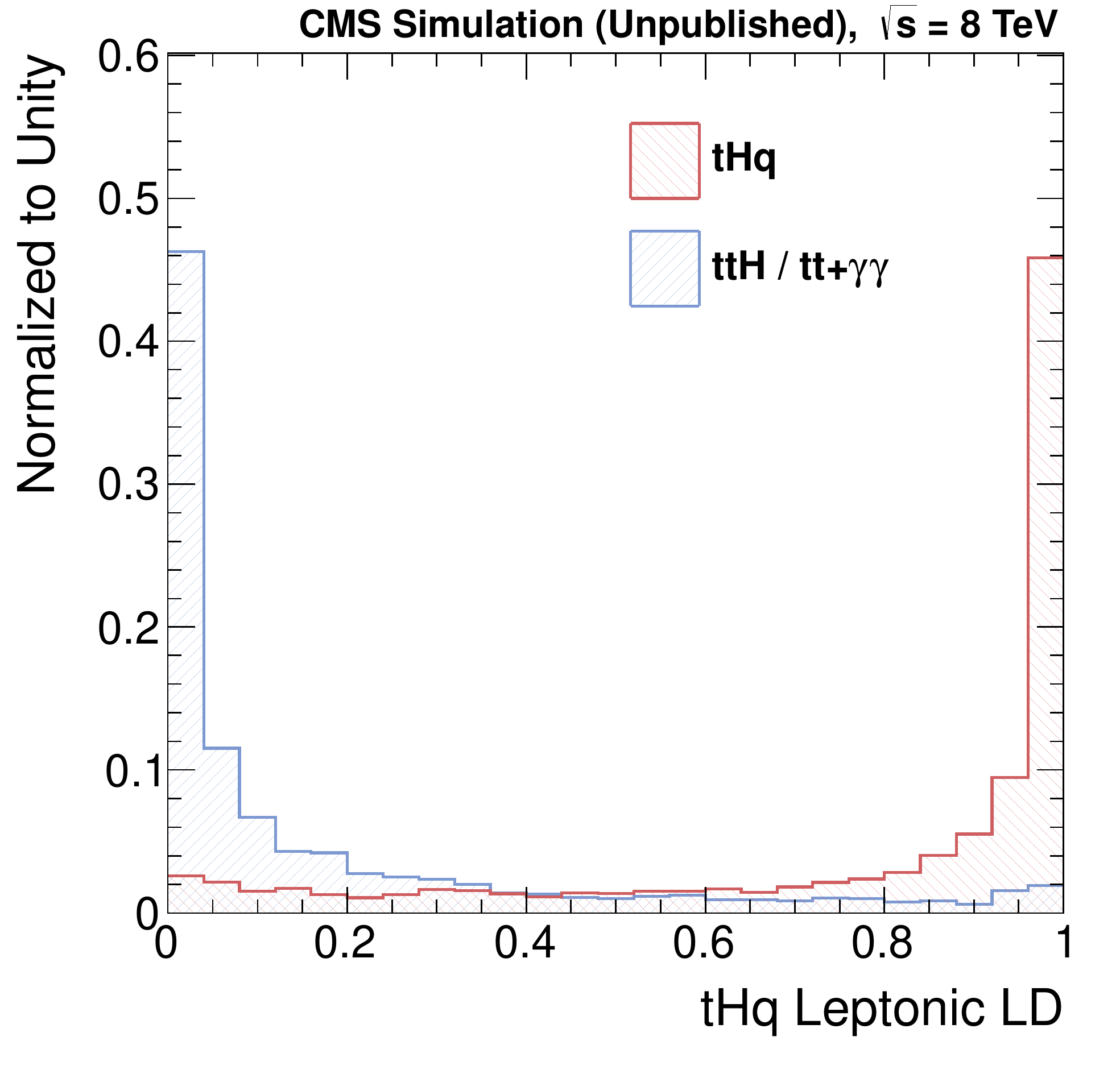}
\caption{Five-variable Likelihood discriminant used to suppress $\ttH$ background. A cut value of LD $>$ 0.25 was optimized to ensure a $\ttH$ contamination below 10\%.}
\label{fig:cmsgg_LD}
\end{minipage}
\hspace{3mm}
\begin{minipage}[t]{0.475\textwidth}
\centering
\includegraphics[width=0.83\textwidth]{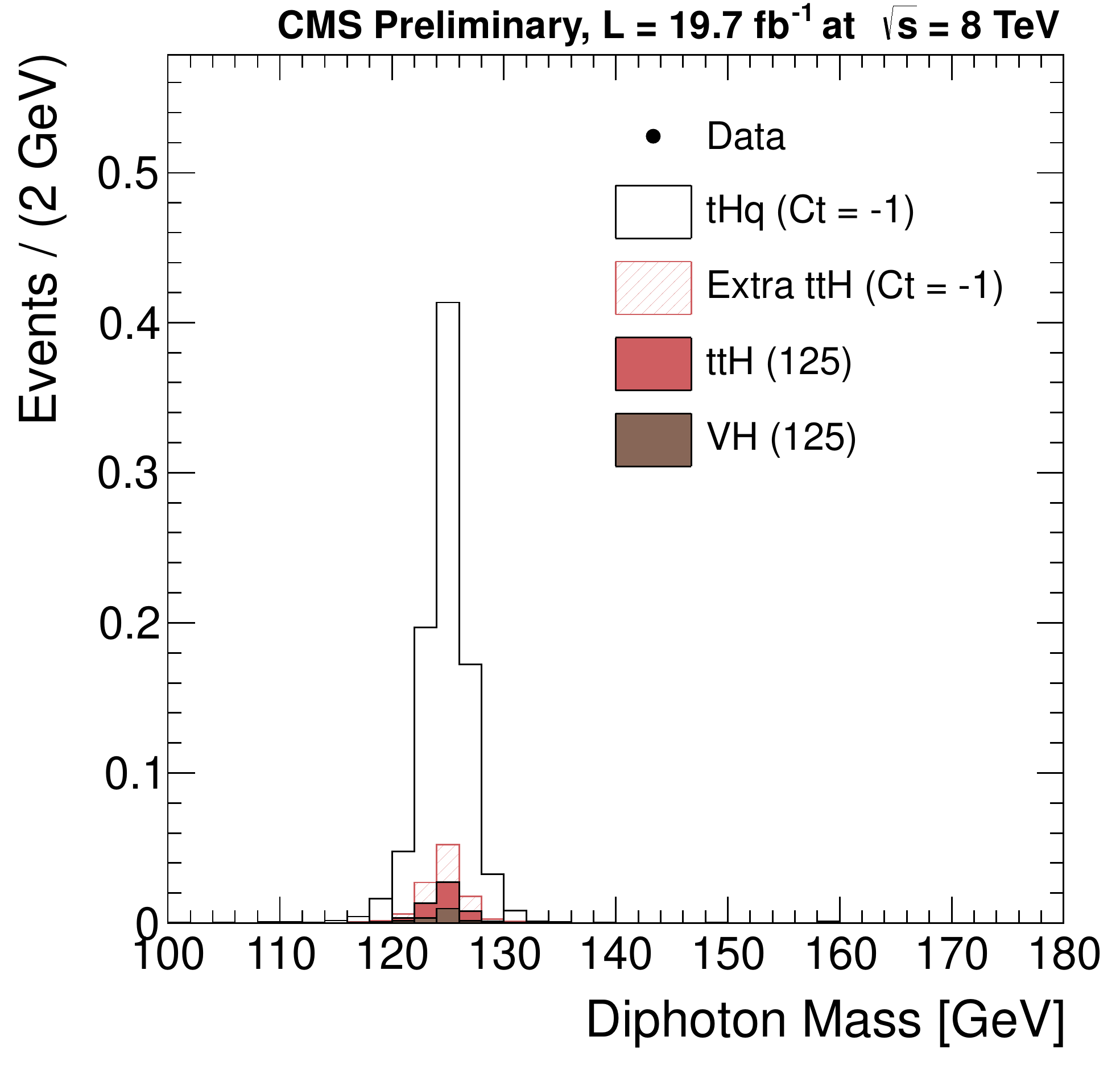}
\caption{Invariant mass of the diphoton system. Zero event of the data (black markers) pass the selection criteria.}
\label{fig:cmsgg_mass}
\end{minipage}
\end{center}
\end{figure}

The analysis described in~\cite{CMS-PAS-HIG-14-001} sets upper limits for anomalous $\tHq$ production with $\Hgg$ decays. As explained above, for the used case $\kt = -1$ both enhancements, in the $\tHq$ cross section and the branching ratio BR(\Hgg), are expected. Like in most diphoton analyses, a resonance is searched for in the diphoton mass $\mgg$ distribution and the signal region is defined as $\pm$ 3 GeV window around the nominal Higgs boson mass.
In the analysis, diphoton events are selected, where the leading photon satisfies $\pT > 50\cdot \mgg/120$ and the subleading photon $\pT > 25$~GeV. Further, an event is asked to possess exactly one lepton (electron~(e) or muon~($\upmu$)) with $\pT > 10$~GeV and a distance $\Delta R > 0.5$ with respect to the photons. At least one b-tagged jet with $\pT > 20$~GeV is required and the hardest additional jet must satisfy $\pT > 20$~GeV and $|\eta| > 1$. 
The reconstruction of the event is straight forward: The Higgs boson candidate is built up from the two photons; the lepton and the b-tagged jet, together with the missing transverse energy are assigned to the top quark candidate and as the light jet candidate the hardest additional jet in the event is taken.
In order to suppress the $\ttH$ background a five-variable likelihood discriminant~(LD) is introduced. The used variables for the discriminant are jet multiplicity, transverse mass of the top quark candidate, pseudorapidity of the light jet candidate, rapidity gap between lepton and light jet and the charge of the lepton candidate. Figure~\ref{fig:cmsgg_LD} shows the performance of the LD and the separation between the $\tHq$ signal and the $\ttH$ background is clearly visible. An additional event requirement of LD~$>$~0.25 ensures a $\ttH$ contamination below 10\%. 
To estimate the signal processes as well as the resonant background events, particularly from VH and $\ttH$ production, Monte Carlo (MC) simulations are used. Other resonant Higgs boson production modes are found to be negligible. The non-resonant backgrounds are estimated via an exponential fit on data in $\mgg$ sideband regions with loosened criteria. Since there are no events in data passing the nominal event selection, the b-tag requirement is slackened or the identification requirement for the photons is inverted. From these values the expected non-resonant background yields in the signal region are extrapolated. 
No events in data are selected in either the signal or the sideband regions as displayed in Figure~\ref{fig:cmsgg_mass}. The observed upper limit coincides with the expected limit of 2.30\,pb, i.e. 4.1 times the predicted cross section in the $\kt=-1$ case. 

\section{Direct search for $\tHq$ and $\Hbb$ with the CMS experiment}
\begin{figure}[hbtp]
\begin{center}
\begin{minipage}[t]{0.475\textwidth}
\centering
\includegraphics[width=\textwidth]{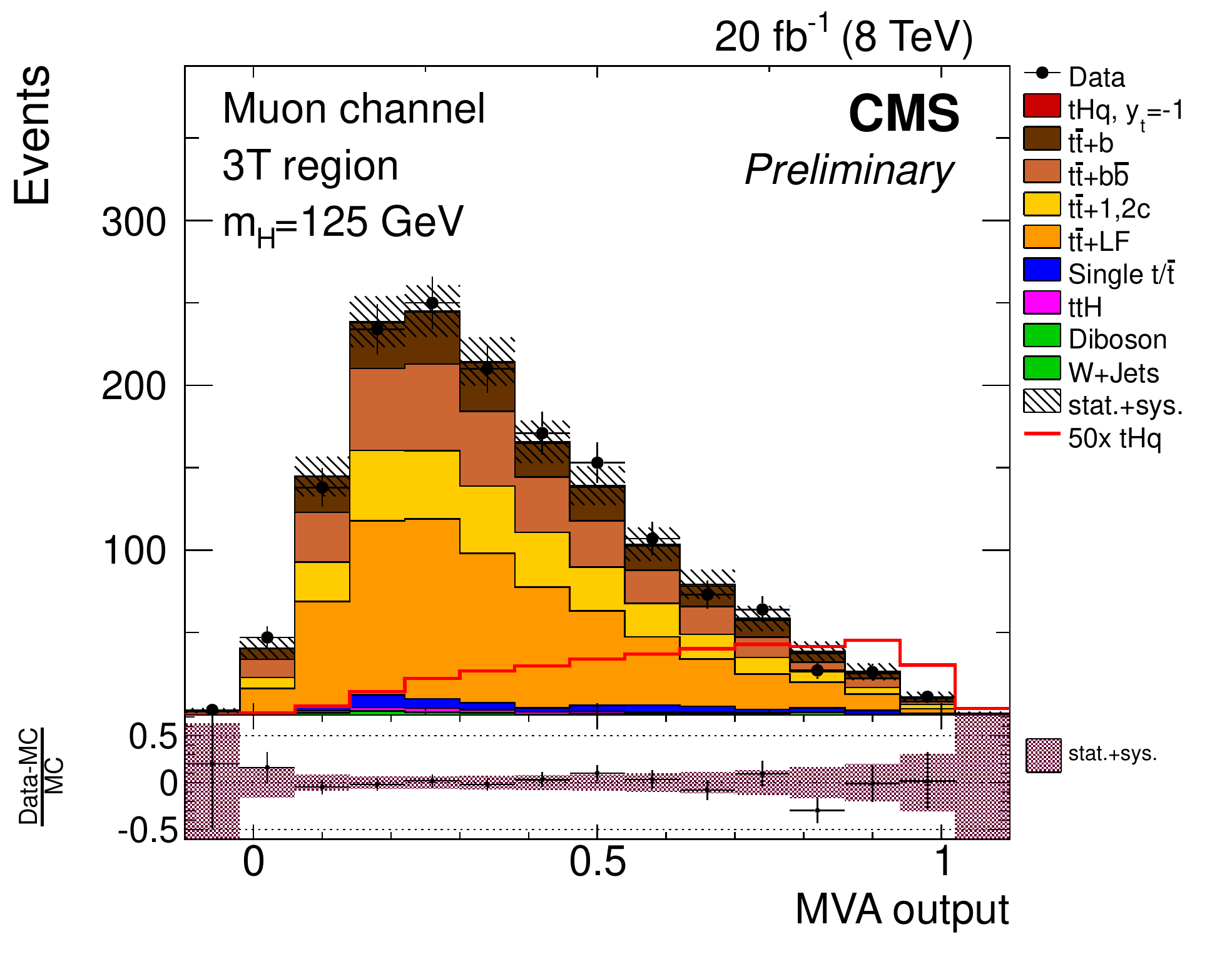}
\caption{\label{fig:3T_mu}Final MVA distribution after the limit extraction fit. Here the data-MC agreement is shown representatively in the muon channel in the 3T region.}
\end{minipage}
\hspace{3mm}
\begin{minipage}[t]{0.475\textwidth}
\centering
\includegraphics[width=0.7841\textwidth]{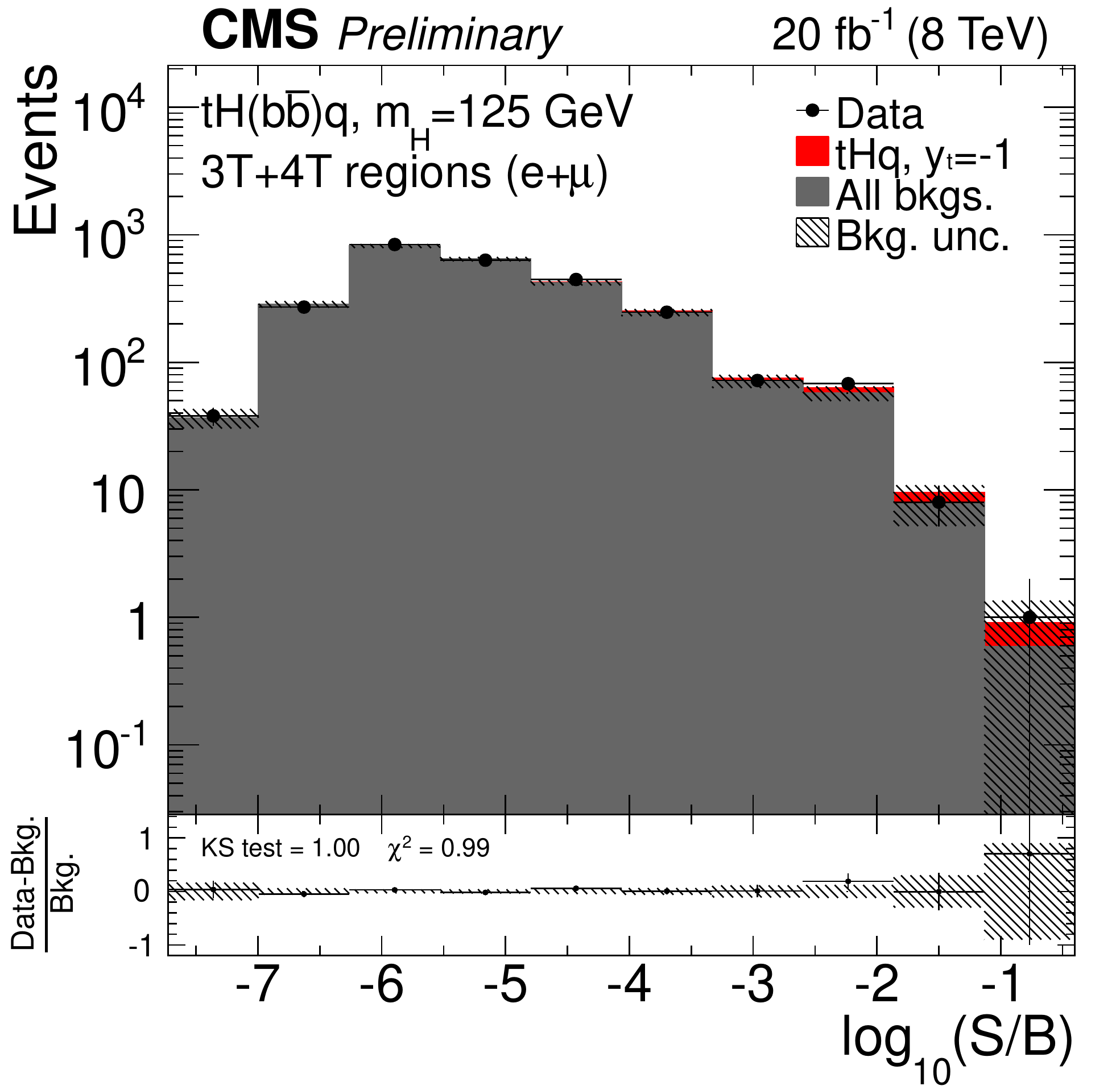}
\caption{\label{fig:SB_comb}Plot containing all events from all signal regions sorted by their expected signal over background ratio. A good agreement between data and background-only expectation is found.} 
\end{minipage} 
\end{center}
\end{figure}

A new analysis optimized for $\kt=-1$ from the CMS collaboration documented in~\cite{CMS-PAS-HIG-14-015} searches also for anomalous $\tHq$ production, but with $\Hbb$ decays in the final state. This challenging multijet signature containing 3 b quarks needs a lot of effort to be sensitive to the signal at all, starting by defining a signal enhanced phase space.
The presence of an additional b quark stemming from initial gluon splitting (qg $\to\tHq^\prime$b) is considered by defining two signal regions asking for either three b-tagged jets (3T region) or for four b-tagged jets (4T region). Additionally at least one untagged jet with $\pT > 20$~GeV is required for every event to account for the forward light jet. Further, the events are asked to have exactly one isolated charged lepton (e or $\upmu$). In the electron channel $\MET > 45$~GeV and in the muon channel $\MET > 35$~GeV is required to suppress QCD background. This signal enhanced phase space still expects a poor signal over background ratio, i.e. $\sim 0.7\%$ in the 3T and $\sim 2.1\%$ in the 4T region. For the latter region only 70 events in data are expected in total making its sensitivity suffer from low statistics. 
The jet assignment to quarks of multijet final states is a combinatorial issue. In $\tHq$ MC simulation for each event a \textit{correct} interpretation is defined as the one with the smallest sum of $\Delta R$ between associated jets and quarks. A multivariate analysis (MVA) using the kinematic information as well as the angular correlations of the event is trained to separate these \textit{correct} interpretations from all other possible \textit{wrong} combinations. When applying the reconstruction to data events all possible combinations are reconstructed and the one with the highest MVA response is chosen. To provide more observables adding sensitivity to the search a similar reconstruction is applied under the assumption it was a $\ttbar$ event. In this way also a handle how background-like the event is can be extracted. Again, characteristic kinematic information and angular correlations are used as inputs for the MVA and from all possible combinations the one with the highest MVA output is taken.
An optimized set out of the observables from both reconstructions together with the lepton charge is chosen as input for the final classification. For this, a third MVA is trained to discriminate $\tHq$ signal events from the background processes $\ttbar$ and $\ttH$. How well the classification MVA performs is shown in Figure~\ref{fig:3T_mu} exemplarily for the 3T muon channel. The simulated signal process in red is clearly separated from the background events, that are dominated by $\ttbar$ production divided into different flavor contents.
Since no significant excess between data and simulation is found, upper limits for anomalous $\tHq$ production are set using a fully frequentist methodology. The observed upper 95\% confidence level limit is 1.77\,pb, i.e. 7.57 times the cross section predicted in the $\kt=-1$ case with $5.14^{+2.14}_{-1.44}$ expected. Given the large uncertainties, the observed limit agrees well with the background-only hypothesis. For a better visualization of these results all events from all channels are sorted by their expected signal over background ratio as shown in Figure~\ref{fig:SB_comb} and the good agreement between data and background is visible.

\section{Constraints on $\kt$ via search for $\ttH$ production with the ATLAS experiment}
\begin{figure}[hbtp]
\begin{center}
\begin{minipage}[t]{0.475\textwidth}
\centering
\includegraphics[width=\textwidth]{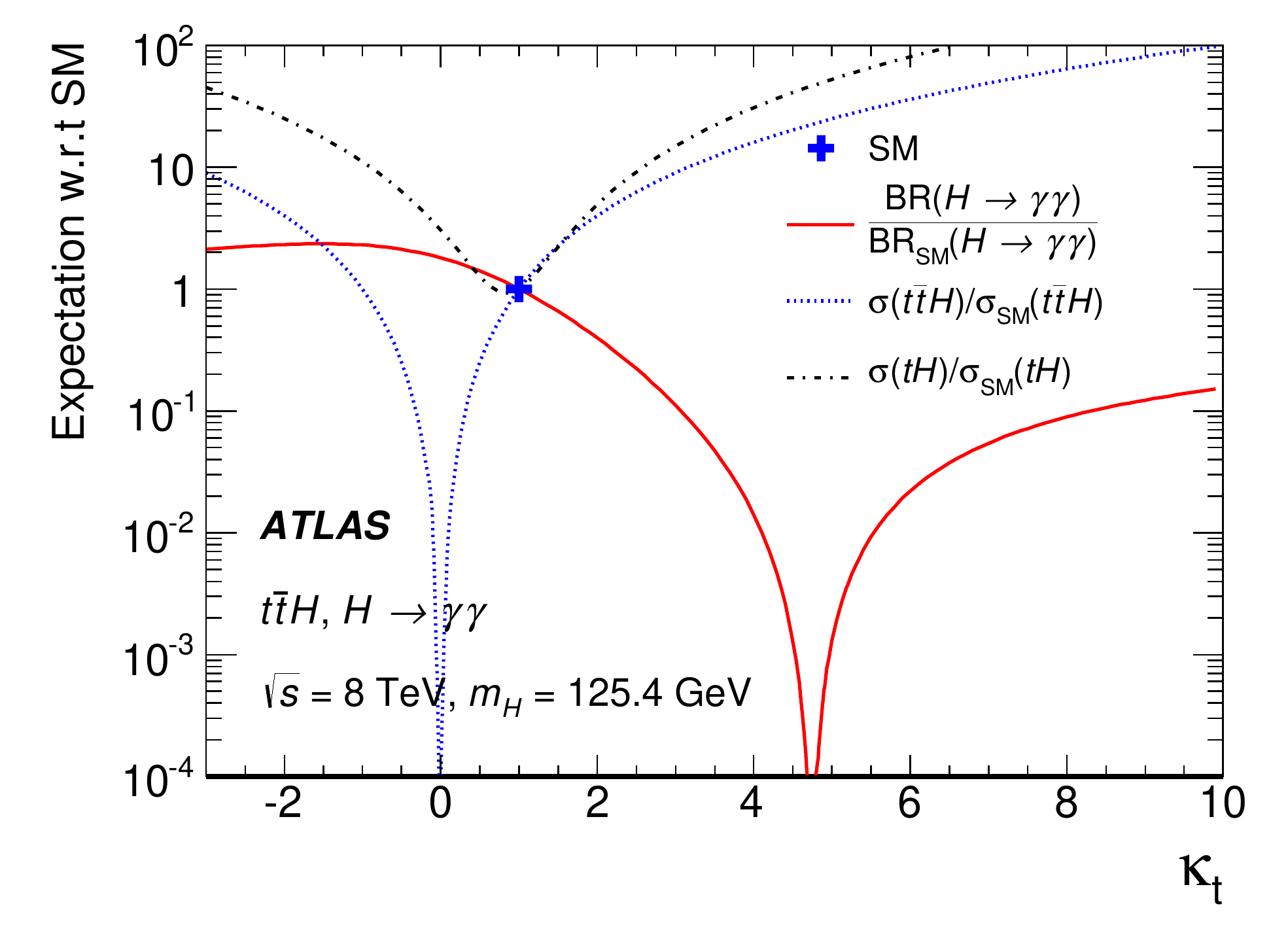}
\caption{$\kt$ dependence of the branching ratio $\Hgg$ and the cross-sections of tH and ttH production relative to the SM expectation ($\kt=0$).}
\label{fig:atlas_xsec}
\end{minipage}
\hspace{3mm}
\begin{minipage}[t]{0.475\textwidth}
\centering
\includegraphics[width=\textwidth]{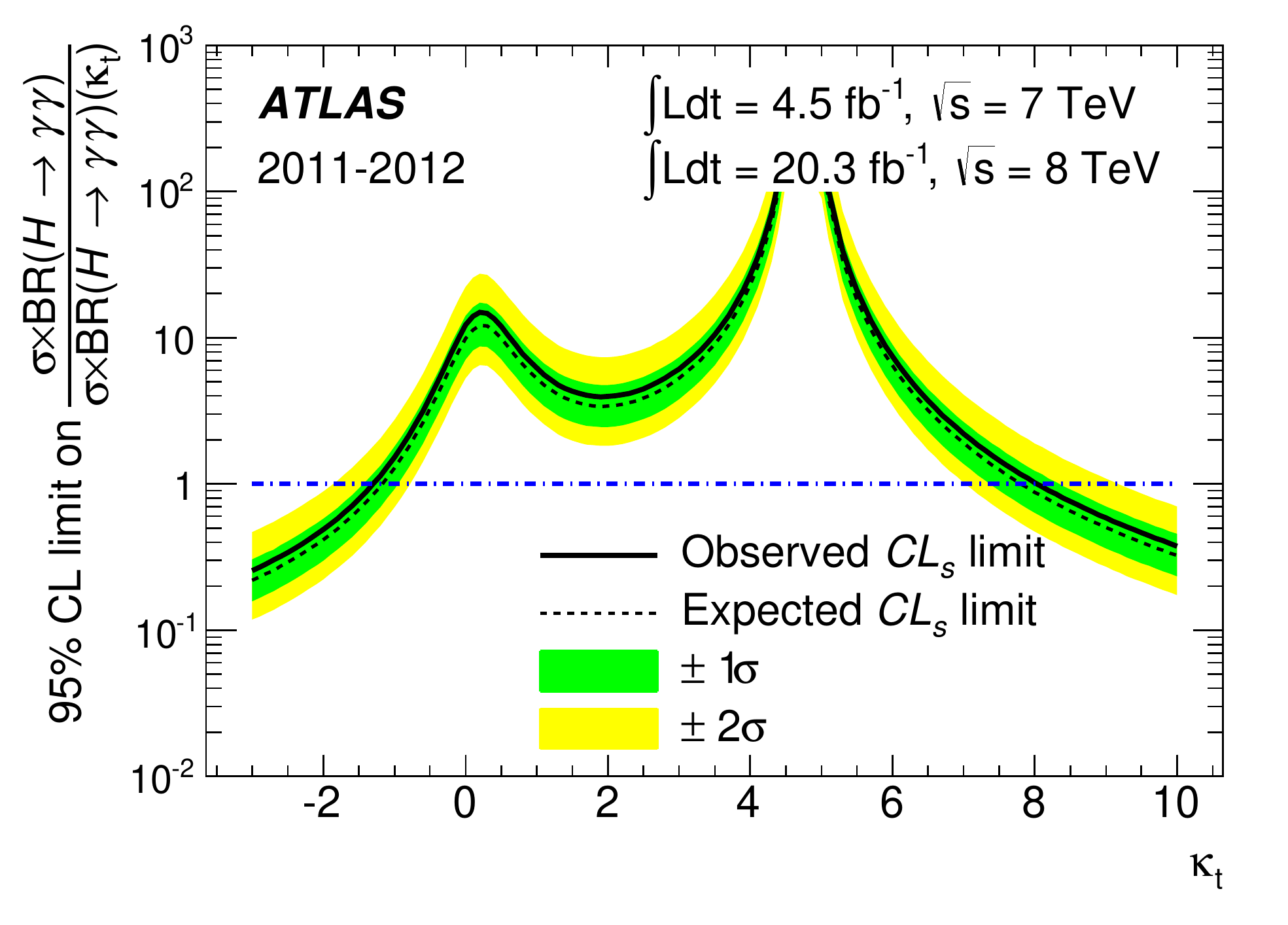} 
\caption{Observed and expected 95\% CL upper limits on the inclusive Higgs production cross section times BR($\Hgg$) with respect to the SM expectation for different $\kt$ values.}
\label{fig:atlas_res}
\end{minipage}
\end{center}
\end{figure}

The ATLAS collaboration investigates the coupling $\kt$ in a different approach documented in~\cite{Aad:2014lma}. Based on the search for $\ttH$ production with $\Hgg$ decays, optimized for the SM case, upper limits get calculated for different $\kt$ values. 
The used events are categorized into two classes, by either requiring zero leptons in the event (hadronic channel) or at least one lepton (leptonic channel). With additional requirements on the number of b-tagged and untagged jets the hadronic channel is optimized for a high purity of $\ttH$ with respect to other Higgs boson production modes, whereas a high $\ttH$ signal efficiency is tried to be achieved in the leptonic channel. In both channels the $\tHq$ production is a non-negligible background. Moreover, the tW-channel associated production of single top quark and Higgs boson (tWH) is considered as well. In the signal region in 8 TeV data ~5\% of all Higgs bosons are expected to stem from $\tHq$ and ~2.4\% from tWH processes, while the dominant fraction comes from $\ttH$ production. 
The fit to extract the limits is performed in the invariant mass distribution of the diphoton system $\mgg$. Combining all channels and both 7 TeV and 8 TeV datasets, the observed upper limit on $\sigma^{\ttH}/\sigma^{\ttH}_{SM}$ is found to be 6.7, with an expected limit of 6.2 when injecting the SM signal as background.
With these results at hand, the dependencies on $\kt$ in a range between -3 and +10 are studied in detail for the affected $\ttH$ and $\tHq$ production cross sections as well as the branching ration BR($\Hgg$). Figure~\ref{fig:atlas_xsec} shows the functions with respect to the SM expectations. The branching ratio BR($\Hgg$) has, due to accidental cancellation of the contributions of top quarks and W bosons to the $\Hgg$ decay, a minimum at +4.7, while the cross section $\sigma^{\ttH}$ is turned off at $\kt=0$.
The results, roughly a reversed reflection of the $\kt$ dependencies, are shown in Figure~\ref{fig:atlas_res}. An upper limit for $\kt$ at 95\% CL of +8.0 and a lower limit of $-1.3$ are observed. 

\section{Conclusion}
The three presented searches show already with the so far available data impressive results. Not only they show a good agreement with the SM, but also make first statements on the tHq production. The CMS collaboration plans to combine all available direct tHq searches, while the ATLAS collaboration wants to add the shown constraints in \cite{ATLAS-CONF-2014-009}.

With the benefit of a higher energy and eventually larger integrated luminosity, the data-taking period starting in 2015 will increase the sensitivity, so at some point a definitive answer on anomalous couplings $\kt$ can be given. 
\section{References}
\input{template.bbl}
\end{document}

%% file: template.bbl
\providecommand{\href}[2]{#2}\begingroup\raggedright\endgroup